\documentclass[dvips,a4paper]{article}
\usepackage{epsfig}
\usepackage{amsmath}

\newcommand{\Lumint}{{\cal L}_{\rm int}}

\renewcommand{\title}[1]{%
  \begin{center}
    \LARGE #1
  \end{center}\par}
\renewcommand{\author}[1]{%
  \vspace{2ex}
  {\Large
   \begin{center}
     \setlength{\baselineskip}{3ex} #1 \par
   \end{center}}}

\begin{document}

\markboth{DVERGSNES, OSLAND, PANKOV and PAVER}
{SEARCH AND IDENTIFICATION OF EXTRA SPATIAL DIMENSIONS AT LHC}

\title{SEARCH AND IDENTIFICATION OF EXTRA SPATIAL DIMENSIONS AT
LHC\footnote{Presented by E.W Dvergsnes at the {\it 6th Alexander Friedmann
International Seminar on Gravitation and Cosmology}, Carg\'ese, 28 June --
3 July, 2004} }

\author{\footnotesize E. W. DVERGSNES$^{a}$\footnote{e-mail:
erik.dvergsnes@ift.uib.no} , 
P. OSLAND$^{a}$,
A. A. PANKOV$^{b}$,
N. PAVER$^{c}$}

\begin{center}
$^{a}$Department of Physics and Technology, University of Bergen, 
All\'{e}gaten 55, N-5007 Bergen, Norway \\
$^{b}$Department of Physics, Pavel Sukhoi Technical University \\
October Avenue 48, 246746 Gomel, Belarus \\
$^{c}$Department of Theoretical Physics, University of Trieste and \\
INFN-Sezione di Trieste
Strada Costiera 11, 34014 Miramare-Grignano (Trieste), Italy 
\end{center}

\begin{abstract}
We present an analysis, based on the center--edge asymmetry, to distinguish
effects of extra dimensions within the Arkani-Hamed--Dimopoulos--Dvali (ADD) 
and Randall--Sundrum (RS) scenarios from other new physics effects in 
lepton-pair production at the CERN Large Hadron Collider LHC. 
Spin-2 and spin-1 exchange can be distinguished 
up to an ADD cutoff scale, $M_H$, of about 5~TeV, at the 95\% CL.
In the RS scenario, spin-2 resonances can be identified in most of
the favored parameter space.
\end{abstract}

\section{Introduction}

A general feature of the different theories extending the Standard Model of
elementary particles (SM) is that new interactions involving heavy elementary
objects and mass scales should exist, and manifest themselves {\it via}
deviations of measured observables from the SM predictions.  While for the
supersymmetric extensions of the SM, there is confidence that the new
particles could be be directly produced and their properties studied, in
numerous other cases, such as the composite models of fermions\cite{Eichten}
and the exchange of leptoquarks,\cite{Buchmuller:1986zs} existing limits
indicate that the heavy states could not be produced even at the highest
energy supercolliders and, correspondingly, only ``virtual'' effects can be
expected. A description of the relevant new interaction in terms of
``effective'' contact-interaction (CI) is most appropriate in these cases. Of
course, since different interactions can give rise to similar deviations from
the SM predictions, the problem is to identify, from a hypothetically measured
deviation, the kind of new dynamics underlying it.

In the context of the hierarchy problem, much attention has been given in the
past few years to the different scenarios involving extra space dimensions and
their manifestations at high energy electron-positron and proton-(anti)proton
colliders. Of particular relevance is the problem of
differentiating their signals from other sources of new phenomena.  We shall
here discuss the possibility of distinguishing such effects of extra
dimensions from other NP scenarios in lepton pair production at the LHC:
\begin{equation}
p+p\to l^+l^-+X,
\label{proc}
\end{equation}
where $l=e,\mu$. Two specific models involving extra dimensions will
be considered, namely the
ADD\cite{Arkani-Hamed:1998rs} and RS\cite{Randall:1999ee} scenarios. 

In the ADD scenario,\cite{Arkani-Hamed:1998rs} gravity is allowed to propagate
in two or more compactified extra space dimensions, with up to millimeter size
$R$. In four dimensions, this mechanism is equivalent to the exchange of a
tower of equally mass-separated Kaluza-Klein (KK) spin-2 states, with
$\Delta M\sim 1/R$.  The relation between the higher-dimensional Planck scale
$M_D$ and the four-dimensional Planck scale $M_{\rm Pl}$ is:
\begin{equation}
M_{\rm Pl}^2\sim R^n\times M_D^{n+2},
\end{equation}
where $n$ is the number of extra dimensions.
The sum over the (almost continuous) spectrum of KK states (of mass 
$m_{\vec n}$) can be expressed as :\cite{Hewett:1999sn}
\begin{equation}
\label{Eq:Hewett}
\sum_{\vec n =1}^\infty \frac{G_{\rm{N}}}{M^2 - m_{\vec n}^2} 
\to \frac{-\lambda}{\pi\,M_H^4},
\end{equation}
where $\lambda$ is a sign factor, $G_{\rm{N}}$ is Newton's constant, and $M_H$
is the cutoff scale, expected to be of the order of the TeV scale.
Equation~(\ref{Eq:Hewett}) can be considered as an effective interaction at
the scale $M_H$.

We will limit ourselves to the simplest version of the RS
scenario,\cite{Randall:1999ee} with only one extra dimension. Differently from
the ADD scenario, there will be narrow graviton spin-2 resonances with
masses of the order of TeV and coupling strength comparable to weak
interactions. Furthermore, the spectrum of KK gravitons in the tower are
unequally spaced, as being located at the Bessel zeros $x_n$:
\begin{equation}
m_n=x_n\,\Lambda_\pi\,\frac{k}{{\bar M}_{\rm Pl}} 
=m_1\,\frac{x_n}{x_1}
\label{Eq:Randall}
\end{equation}
where $\Lambda_\pi$ is the KK coupling strength.

This model has two independent parameters, conveniently taken to be $k/\bar
M_{\rm{Pl}}$ and $m_1$, where $k$ is a constant of ${\cal O}(\bar
M_{\rm{Pl}})$, and $m_1$ is the mass of the first graviton resonance.  Also,
the phenomenology is quite different from that of the ADD scenario, in the
sense that RS resonances may well be in the energy range of the LHC, and hence
show up as peaks in the cross section.

\section{Center--edge asymmetry $A_{\rm{CE}}$}

In the SM, lepton pairs can at hadron colliders be produced at tree-level 
via the following parton-level process
\begin{equation}\label{Eq:qqbar-SM}
q \bar q \to \gamma,Z \to l^+ l^-.
\end{equation}
Now, if gravity can propagate in extra dimensions, 
the possibility of KK graviton 
exchange opens up two tree-level channels in addition to 
the SM channels, namely
\begin{equation} \label{Eq:qqbar-gg}
q \bar q \to G \to l^+ l^-, \qquad \text{and} \qquad
gg \to G \to l^+ l^-,
\end{equation}
where $G$ represents the gravitons of the KK tower.

The center--edge and total cross sections can at the parton level be defined
like for initial-state electrons and 
positrons:\cite{Osland:2003fn}
\begin{equation}
\label{Eq:sigma-hat-ce}
\hat \sigma_{\rm{CE}}
\equiv \left[\int_{-z^*}^{z^*} 
- \left(\int_{-1}^{-z^*}
+\int_{z^*}^{1}\right)\right]
\frac{d\hat \sigma}{dz}\, dz, \quad
\hat \sigma
\equiv \int_{-1}^{1} 
\frac{d\hat \sigma}{dz}\, dz,
\end{equation}
where $z=\cos\theta_{\rm{cm}}$, with $\theta_{\rm{cm}}$ the angle, in the
c.m.\ frame of the two leptons, between the lepton and the proton. Here,
$0<z^*<1$ is a parameter which defines the border between the ``center'' and
the ``edge'' regions.  This asymmetry has been demonstrated very selective to
the ADD effects in the electron-positron case,\cite{Osland:2003fn} and we want
to test its use in the more complicated (but experimentally forthcoming)
subprocesses (\ref{Eq:qqbar-SM}) and (\ref{Eq:qqbar-gg}).

The center--edge asymmetry can then
for a given dilepton invariant mass $M$ be defined as 
\begin{equation}
\label{Eq:ace}
A_{\rm{CE}}(M)=\frac{d\sigma_{\rm{CE}}/dM}
                 {d\sigma/dM},
\end{equation}
where a convolution over parton momenta is performed,
and we obtain $d\sigma_{\rm{CE}}/dM$ and $d\sigma/dM$ from 
the inclusive differential cross sections $d\sigma_{\rm{CE}}/dM\,dy\,dz$ and
$d\sigma/dM\,dy\,dz$, respectively, by integrating over $z$ according
to Eq.~(\ref{Eq:sigma-hat-ce}) and over rapidity $y$ between
$-Y$ and $Y$,  with $Y=\log(\sqrt{s}/M)$.\cite{Dvergsnes:2004}

For the SM contribution to the center--edge asymmetry,
the convolution integrals, depending on the parton distribution functions,
cancel, and one finds\cite{Dvergsnes:2004}
\begin{equation}
\label{Eq:ACEspin1}
A_{\rm{CE}}^{\rm{SM}}
=\frac{1}{2} z^*(z^{*2}+3)-1.
\end{equation} 
This result is thus independent of the dilepton mass $M$, and identical to the
result for $e^+e^-$ colliders.\cite{Osland:2003fn} Hence, in the case of no
cuts on the angular integration, there is a unique value, $z^*=z_0^*\simeq
0.596$, for which $A_{\rm{CE}}^{\rm{SM}}$ vanishes, corresponding to
$\theta_{\rm{cm}} = 53.4^\circ$.

The SM center-edge asymmetry of Eq.~(\ref{Eq:ACEspin1}) is equally valid for a
wide variety of NP models: composite-like contact interactions, $Z'$ models,
TeV-scale gauge bosons, {\it etc}. However, if graviton exchange is possible,
the graviton tensor couplings would yield a different angular distribution,
leading to a different dependence of $A_{\rm{CE}}$ on $z^*$. In this case, the
center--edge asymmetry would not vanish for the above choice of $z^* = z_0^*$.
Furthermore, it would show a non-trivial dependence on $M$. Thus, a value for
$A_{\rm{CE}}$ different from $A_{\rm{CE}}^{\rm{SM}}$ would indicate
non-vector-exchange NP.

Another important difference from the SM case is that the graviton also
couples to gluons, and therefore it has the additional $gg$ initial state of
Eq.~(\ref{Eq:qqbar-gg}) available. In summary then, including graviton
exchange and also experimental cuts relevant to the LHC detectors, the
center--edge asymmetry is no longer the simple function of $z^*$ given by
Eq.~(\ref{Eq:ACEspin1}).\cite{Dvergsnes:2004}

\section{Identifying graviton exchange and graviton resonance}
We assume now that a deviation from the SM is discovered in the
cross section, either in the form of a CI or a resonance. We
will here investigate in which regions of the ADD and RS parameter spaces such
a deviation can be {\it identified} as being caused by spin-2 exchange.  More
precisely, we will see how the center--edge asymmetry (\ref{Eq:ace}) can be 
used to exclude spin-1 exchange interactions beyond that of the SM. At the 
LHC, with luminosity ${\cal L}_{\rm int}=100$ and $300~{\rm{fb}}^{-1}$, we 
require the invariant lepton mass $M>400$~GeV and 
divide the data into $200$~GeV bins as long as the number of events in each 
bin, $\epsilon_l {\cal L}_{\rm{int}}\sigma(i)$, is larger than 10. Here, 
$\epsilon_l$ is the experimental reconstruction efficiency and 
$\sigma (i)$ the cross section in bin $i$.   
To compute cross sections we use the CTEQ6 parton 
distributions.\cite{Pumplin:2002vw} We impose angular cuts relevant to the 
LHC detectors. The lepton pseudorapidity cut is $|\eta|<\eta_{\rm{cut}}=2.5$
for both leptons, and in addition to the angular cuts, we impose on each lepton
a transverse momentum cut $p_\perp>p_\perp^{\rm{cut}}=20~{\rm GeV}$.

From a conventional $\chi^2$ analysis we find the ADD-scenario {\it
identification} reach on $M_{H}$ at the LHC summarized in Table~1.  In this
table we also include the identification reach obtained from the analysis of
the center-edge asymmetry performed at an $e^+e^-$ Linear Collider (LC) for
c.m. energy 500~GeV.
\vspace*{-10pt}
\begin{table}[h]
\caption{Identification reach on $M_H$ at 95\% CL from $A_{\rm{CE}}$.}
{\begin{tabular}{lcccc}
\hline
Collider & LHC $100~{\rm{fb}}^{-1}$ &  LHC $300~{\rm{fb}}^{-1}$ 
&  LC $50~{\rm{fb}}^{-1}$ & LC $500~{\rm{fb}}^{-1}$ \\
\hline
$\lambda=+1$ (TeV) & 4.8 &  5.4 & 3.1 & 4.1 \\
$\lambda=-1$ (TeV) & 5.0 &  5.9 & 3.1 & 4.1 \\ \hline
\end{tabular}}
\end{table}
\vspace*{-10pt}

As displayed in Eq.~(\ref{Eq:Randall}), in the RS scenario the resonances 
are unevenly spaced. If the first resonance is sufficiently heavy, the 
second resonance would be difficult to resolve
within the kinematical range allowed experimentally at the LHC, and we 
shall now consider this situation. 

We choose a 200~GeV bin around the RS resonance mass $m_1$, and obtain the
results presented in Fig.~1, where we display the 2, 3 and $5\sigma$ contours
for $\Lumint=100$ and $300~\text{fb}^{-1}$.
As shown in this figure, the identification reach at the LHC provided by the
observable $A_{\rm CE}$ covers a large portion of the ``theoretically
preferred'' (in order not to create additional hierarchies) parameter space
$\Lambda_\pi<{\cal O}(10)$ TeV. For $k/\bar M_{\rm{Pl}}=0.1$, the
$\Lumint=100~\text{fb}^{-1}$ identification reach extends above $m_1\simeq
3.5~{\rm TeV}$ (at the $2\sigma$ level).
\begin{figure}[ht]
\begin{center}
\setlength{\unitlength}{1cm}
\begin{picture}(12,8)
\put(1.5,0.0)
{\mbox{\epsfysize=9cm\epsffile{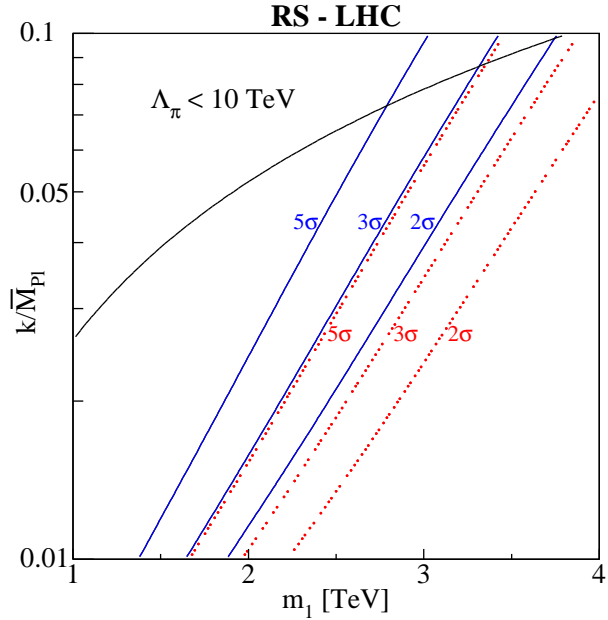}}}
\end{picture}
\caption{Spin-2 identification of an RS resonance, using the center--edge
asymmetry, integrated over bins of 200~GeV around the peak. Solid (dotted)
$2\sigma$, $3\sigma$, $5\sigma$ contours: $\Lumint=100~\text{fb}^{-1}$
($300~\text{fb}^{-1}$).  The theoretically favored region,
$\Lambda_\pi<10~{\rm{TeV}}$, is indicated.}
\end{center}
\end{figure}

In conclusion, we have considered the ADD scenario parametrized by $M_H$, and
the RS scenario parametrized by $m_1$ and $k/\bar M_{\rm{Pl}}$.  Although
somewhat higher sensitivity reaches on $M_H$ or $m_1$ than obtained here are
given by other 
approaches, this method based on $A_{\rm{CE}}$ is suitable for actually
{\it pinning down} the spin-2 nature of the KK gravitons up to very high 
$M_H$ or $m_1$. This is different from just detecting deviations from the 
Standard Model predictions, and is a way to obtain additional information 
on the underlying new-physics scenario and to impose stringent constraints 
on the extra dimension scenarios here discussed. Therefore, the analysis 
sketched here can potentially represent a valuable method complementary to 
the direct fit to the angular distribution of the lepton 
pairs.\cite{Allanach:2000nr,Davoudiasl:2000wi} 
Also, the analysis can readily be extended to other final states, in high 
energy proton-proton collisions, different from $l^+l^-$ in (\ref{proc}), 
such as di-photon or di-jet final states.


\begin{thebibliography}{0}

\bibitem{Eichten} E. Eichten, K.Lane and M. E. Peskin, Phys. Rev. Lett. 
{\bf 50}, 811 (1983); R.R\"uckl, Phys. Lett. B {\bf 129}, 363 (1983).

\bibitem{Buchmuller:1986zs}
W.~Buchmuller, R.~Ruckl and D.~Wyler,
Phys.\ Lett.\ B {\bf 191}, 442 (1987)
[Erratum-ibid.\ B {\bf 448}, 320 (1999)].


\bibitem{Arkani-Hamed:1998rs}
N.~Arkani-Hamed, S.~Dimopoulos and G.~R.~Dvali,
Phys.\ Lett.\ B {\bf 429}, 263 (1998);
%
I.~Antoniadis, N.~Arkani-Hamed, S.~Dimopoulos and G.~R.~Dvali,
Phys.\ Lett.\ B {\bf 436}, 257 (1998)
[arXiv:hep-ph/9804398].


\bibitem{Randall:1999ee}
L.~Randall and R.~Sundrum,
Phys.\ Rev.\ Lett.\  {\bf 83}, 3370 (1999).


\bibitem{Hewett:1999sn}
J.~L.~Hewett,
Phys.\ Rev.\ Lett.\  {\bf 82}, 4765 (1999).

\bibitem{Osland:2003fn}
P.~Osland, A.~A.~Pankov and N.~Paver,
Phys.\ Rev.\ D {\bf 68}, 015007 (2003).

\bibitem{Dvergsnes:2004}
For details of the analysis and original references, see
E.W.~Dvergsnes, P.~Osland, A.~A.~Pankov and N.~Paver,
Phys.\ Rev.\ D {\bf 69}, 115001 (2004).

\bibitem{Pumplin:2002vw}
J.~Pumplin, D.~R.~Stump, J.~Huston, H.~L.~Lai, P.~Nadolsky and W.~K.~Tung,
JHEP {\bf 0207}, 012 (2002). 

\bibitem{Allanach:2000nr}
B.~C.~Allanach, K.~Odagiri, M.~A.~Parker and B.~R.~Webber,
JHEP {\bf 0009}, 019 (2000).

\bibitem{Davoudiasl:2000wi}
H.~Davoudiasl, J.~L.~Hewett and T.~G.~Rizzo,
Phys.\ Rev.\ D {\bf 63}, 075004 (2001). 

\end{thebibliography}
\end{document}